\documentclass[preprint,showpacs,preprintnumbers,amsmath,amssymb,natbib]{revtex4}

\usepackage{graphicx}
\usepackage{epsfig}		
\usepackage{dcolumn}
\usepackage{bm}
\def\0{\mbox{\tiny $0$}}
\def\1{\mbox{\tiny $1$}}
\def\2{\mbox{\tiny $2$}}
\def\3{\mbox{\tiny $3$}}
\def\4{\mbox{\tiny $4$}}
\def\5{\mbox{\tiny $5$}}
\def\6{\mbox{\tiny $6$}}
\def\7{\mbox{\tiny $7$}}
\def\8{\mbox{\tiny $8$}}
\def\9{\mbox{\tiny $9$}}

\def\f14{\mbox{\tiny $\frac{1}{4}$}}

\def\R{\mbox{\tiny $R$}}

\def\B{\mbox{\tiny $B$}}

\def\mi{\mbox{\tiny $-$}}

\def\bb#1{\mbox{\footnotesize $(#1)$}}

\begin{document}

\title{Thermodynamic equilibrium conditions for mass varying particle structures}

\author{A. E. Bernardini}
\affiliation{Departamento de F\'{\i}sica, Universidade Federal de S\~ao Carlos, PO Box 676, 13565-905, S\~ao Carlos, SP, Brasil}
\email{alexeb@ufscar.br, alexeb@ifi.unicamp.br}
\author{O. Bertolami}
\affiliation{Instituto Superior T\'ecnico, Departamento de F\'{\i}sica, Av. Rovisco Pais, 1, 1049-001, Lisboa, Portugal}
\email{orfeu@cosmos.ist.utl.pt}
\altaffiliation[Also at]{~Instituto de Plasmas e Fus\~{a}o Nuclear, IST, Lisbon}

\date{\today}

\begin{abstract}
The thermodynamic equilibrium conditions for compact structures composed by mass varying particles are discussed assuming that the so-called dynamical mass behaves like an additional extensive thermodynamic degree of freedom.
It then follows that, such weakly interacting massive particles can form clusters of non-baryonic matter or even astrophysical objects that, at cosmological scales, are held together by gravity and by the attractive force mediated by a background scalar field.
The equilibrium conditions for resultant static, spherically symmetric objects of uniform energy density are derived for the case where the particles share essential features with the mass varying mechanism in the context of cosmological scenarios.
Physical solutions which result in stable astrophysical lumps are obtained and discussed.
\end{abstract}

\date{\today}
\maketitle

The understanding of the theory of stellar evolution requires the knowledge of all interactions among the particles that make up the astrophysical objects.
Regardless the specific nature of these interactions, a common property is the additivity of the energy for a macroscopic system: if the system is divided into macroscopic parts, then the interaction energy between these parts will be negligibly small and for short-range forces one can introduce the concept of specific energy $\epsilon = E/N = \rho/n$, where $\rho$ is the energy density ($E$ is the total energy in a volume $V$), and $n$ is the density of particles to which one relates the energy ($N$ is the total number of particles, $N = n\, V$).
In opposition, when the interaction is important, the pressure, $p$, allows one to define the interaction force between parts of a system and it depends only on the state of the matter at a contact surface: $p = p(n,\,s)$, i. e. $p$ depends on the density of particles $n$ and on the specific entropy $s = S/N$.
The other two relevant intensive thermodynamic variables are the temperature $T(n,\,s)$ and the chemical potential $\mu(n,\,s)$.
These quantities can be put together in the macroscopic version of the first law of thermodynamics,
\begin{equation}
\mbox{d}E = - p(V,\,S)\,\mbox{d}V + T(V,\,S)\,\mbox{d}S + \sum_j \mu_j \,\mbox{d}N_j
\label{star01}
\end{equation}
where $V$ is the volume of confinement, and the indices $j$ correspond to different types of particles.

Assuming a fluid element $V$ whose moving walls are attached to the fluid so that no particles flow in or out, the total number of particles $N =\sum_j N_j$ in it must remain fixed.
For neutral matter and for only one-type particle fluid, it results in a particle number conservation law,
\begin{equation}
\mbox{d}N = \mbox{d}(n\, V) = 0.
\label{star03}
\end{equation}
In this case, the volume occupied by one particle can be expressed by $V = 1/n$, its corresponding entropy by $S = N\,s = n\,V\,s = s$, and the above thermodynamic relation (\ref{star01}) by
\begin{equation}
\mbox{d}\epsilon = \mbox{d}\left(\frac{\rho}{n}\right) =  - p(n,\,s)\,\mbox{d}\left(\frac{1}{n}\right) + T(n,\,s)\,\mbox{d}s.
\label{star04}
\end{equation}
Thus,
\begin{eqnarray}
\mbox{d}\rho &=& \left(\frac{\partial\rho}{\partial n}\right)_s \mbox{d}n + \left(\frac{\partial\rho}{\partial s}\right)_n \mbox{d}s\nonumber\\
&=&\frac{\rho(n,\,s) + p(n,\,s)}{n} \mbox{d}n + n\,T(n,\,s)\,\mbox{d}s,
\label{star05}
\end{eqnarray}
and the analysis is further simplified when the one-type particle fluid is everywhere endowed with the same entropy per particle, $s$.
The pressure and energy density are then related by
\begin{equation}
n \left(\frac{\partial\rho}{\partial n}\right)_s  = \rho(n,\,s) + p(n,\,s).
\label{star06}
\end{equation}

The mass varying mechanism considers the possibility of particle masses arising from an interaction with an underlying scalar field which describes the dynamics of the dark energy component of the universe.
At our approach, we assume that the such a dynamical mass behaves like an additional extensive thermodynamic degree of freedom.
The mass varying behaviour is achieved from the analytical dependence of the scalar field on the radial coordinate of a curved space-time, in analogy with the well-established cosmological mass varying mechanism \cite{Hun00,Gu03,Far04,Pec05,Bro06A,Bja08,Ber08A,Ber08B}.
The mutual interaction among neutral particles in stellar conditions is driven by the coupling to a cosmological scalar field dark energy component through a dynamical mass.
In particular, one finds that the modified equilibrium conditions for static, spherically symmetric objects of uniform energy density share some key features with the mass varying mechanism in cosmology scenarios.
The resulting astrophysical objects correspond to lumps or compact structures of mass varying non-baryonic matter held together by gravity and the attractive force mediated by the background scalar field.
Our analysis reveals that such static solutions become dynamically unstable for different Buchdahl's limits \cite{Buc59} for the ratio $M/R$ between total mass-energy, $M$, and stellar radius, $R$.

For cosmological scenarios that admit this kind of mass varying mechanism, the instabilities signalled by an imaginary speed of sound in the non-relativistic regime may play a crucial role in the structure formation.
In previous discussions, it has been pointed out that mass varying models generically face stability problems for some choices of scalar field couplings and scalar field potentials once the mass varying particle becomes non-relativistic \cite{Bea08,Bro06A,Pas06,Wet02}.
The natural interpretation of these instabilities is that the Universe becomes inhomogeneous with these overdensities which are subject to nonlinear fluctuations that eventually collapse into lumps or clusters.

Presumably, the neutrino with mass $m_{\nu}$ is an example of non-baryonic matter and has its origin on the vacuum expectation value (VEV) of a scalar field.
Naturally the dynamical mass behaviour is governed by the dependence of the scalar field on space-time coordinates like the cosmological scale factor, $a$.

Given a particle statistical distribution $f\bb{q}$, where $q \equiv \frac{|\mbox{\boldmath$p$}|}{T_{\0}}$, $T_{\0}$ being the background temperature at present, in the flat FRW cosmological scenario, the corresponding energy density and pressure can be expressed by
\begin{eqnarray}
\rho\bb{a, \phi} &=&\frac{T^{\4}_{\nu \0}}{\pi^{\2}\,a^{\4}}
\int_{_{0}}^{^{\infty}}{\hspace{-0.3cm}dq\,q^{\2}\, \left(q^{\2}+\frac{m^{\2}\bb{\phi}\,a^{\2}}{T^{\2}_{\0}}\right)^{\1/\2}\hspace{-0.1cm}f\bb{q}},\\
p\bb{a, \phi} &=&\frac{T^{\4}_{\0}}{3\pi^{\2}\,a^{\4}}\int_{_{0}}^{^{\infty}}{\hspace{-0.3cm}dq\,q^{\4}\, \left(q^{\2}+\frac{m^{\2}\bb{\phi}\,a^{\2}}{T^{\2}_{\nu \0}}\right)^{\mi\1/\2}\hspace{-0.1cm} f\bb{q}},~~~~ \nonumber
\label{gcg01}
\end{eqnarray}
where sub-index $0$ denotes present-day values, for which $a_{\0} = 1$.
A simple mathematical manipulation allows one to show that
\begin{equation}
m\bb{\phi} \frac{\partial \rho\bb{a, \phi}}{\partial m\bb{\phi}} = \rho\bb{a, \phi} - 3 p\bb{a, \phi},
\label{gcg02}
\end{equation}
and, from the dependence of $\rho$ on $a$, one obtains the energy-momentum conservation for the fluid,
\begin{equation}
\dot{\rho}\bb{a, \phi} + 3 H (\rho\bb{a, \phi} + p\bb{a, \phi}) =
\dot{\phi}\frac{d m\bb{\phi}}{d \phi} \frac{\partial \rho\bb{a, \phi}}{\partial m\bb{\phi}},
\label{gcg03}
\end{equation}
where $H = \dot{a}/{a}$ is the expansion rate of the Universe.
Thus the dependence of $m$ on $\phi$ turns it into a dynamical behaviour.

Analyzing structure formation, through Eq.~(\ref{gcg02}), the mass of these particles depends on the value of a slowly varying classical scalar field \cite{Ber08A,Ber08B,Wet94,Bea08,Wet02}.
For classical particles, the action of the field coupled to mass varying particle system can be written as \cite{Wet08,Tet08}
\begin{equation}
S = \int{\mbox{d}^{\4} x \sqrt{-g} \left[\frac{R}{16 \pi G} + \frac{1}{2}g^{\mu\nu}\partial_{\mu}\phi \partial_{\nu}\phi + V\bb{\phi} + \rho\bb{\phi}\right]}
\label{gcg03B}
\end{equation}
where $G$ is the Newton constant, $R$ is the Ricci curvature and $\rho\bb{\phi}$ is the corresponding mass varying particle density.
For static, spherically symmetric solutions, one employs the Schwarzschild metric given by $\mbox{d}s^{\2} = -B\bb{r} \mbox{d}t^{\2} + A\bb{r} \mbox{d}r^{\2} + r^{\2}\left(\mbox{d}\theta^{\2} + \sin^{\2}(\theta)\mbox{d}\varphi^{\2}\right)$, where $A = B^{-\1} = (1 - 2 M /r)^{-\1}$, with $G = 1$, and $r$ is the radial coordinate.
And from the field equations, one can write
\begin{eqnarray}
\frac{\mbox{d}^{2}\phi}{\mbox{d} r^{2}} + \left(\frac{2}{r} + \frac{B^{\prime}}{2 B} - \frac{A^{\prime}}{2 A}\right)
\frac{\mbox{d}\phi}{\mbox{d} r}
&=&
A \left (\frac{\mbox{d} V\bb{\phi}}{\mbox{d} \phi}  + \frac{\partial \rho\bb{\phi}}{\partial \phi} \right)\nonumber\\
&=&
A \left (\frac{\mbox{d} V\bb{\phi}}{d \phi}  + \frac{\mbox{d} \ln{m\bb{\phi}}}{d \phi} (\rho - 3 p)\right)
\label{gcg03C}
\end{eqnarray}
from which the explicit dependence of $\phi$ on $r$ is obtained.

Assuming the adiabatic approximation at cosmological scales \cite{Bea08}, the cosmological stationary condition \cite{Far04,Bro06A,Bja08,Ber08A,Ber08B} could be applied to Eq.~(\ref{gcg03C}) in order to suppress its right hand side.
As a natural consequence of the solutions of Eq.~(\ref{gcg03C}), the analytical dependence of the scalar field on $r$ is suppressed as one is away from the concentration of matter, i. e. for large values of the radial coordinate.
In this case, one recovers the flat FRW background Universe.
By adding the usual cosmological time dependent terms, Eq.~(\ref{gcg03C}) should be reduced to the scalar field equation of motion,
\begin{equation}
\ddot{\phi} + 3 H \dot{\phi} + \left (\frac{\mbox{d} V\bb{\phi}}{d \phi}  + \frac{\mbox{d} \ln{m\bb{\phi}}}{d \phi} (\rho - 3 p)\right)
= 0,
\label{gcg08}
\end{equation}
and, independently of the dependence of $\phi$ on space-time coordinates ($a$ or $r$), the stationary condition should remain valid in the adiabatic regime, i. e.
\begin{equation}
\frac{\mbox{d} V\bb{\phi}}{d \phi}  + \frac{\mbox{d} \ln{m\bb{\phi}}}{d \phi} (\rho - 3 p)
= 0.
\label{gcg08C}
\end{equation}
Thus, the simplest scenario is obtained from the extension of the cosmological stationary condition to the static configuration which leads to Eq.~(\ref{gcg03C}).
Introducing the abovementioned Schwarzschild expressions for $A\bb{r}$ and $B\bb{r}$, and assuming the stationary condition, Eq.~(\ref{gcg03C}) can be simplified as
\begin{eqnarray}
\frac{\mbox{d}^{2}\phi}{\mbox{d} r^{2}} + \left[\frac{2 - 8(M/R)(r^{\2}/R^{\2})}{r - 2(M/R) (r^{\3}/R^{\2})}\right]
\frac{\mbox{d}\phi}{\mbox{d} r}
&=& 0 ~~~~ (r < R),\nonumber\\
\frac{\mbox{d}^{2}\phi}{\mbox{d} r^{2}} + \left(\frac{2(r - M)}{r - 2 M}\right)
\frac{\mbox{d}\phi}{\mbox{d} r}
&=& 0 ~~~~ (r > R),
\label{gcg03CDM}
\end{eqnarray}
which gives
\begin{eqnarray}
\phi\bb{r} &=& \phi^{in}_{\0} + \frac{\phi^{in}_{\1}}{2M} \ln{\left(1 - \frac{2M}{r}\right)} ~~~~ (r < R),\nonumber\\
\phi\bb{r} &=& \phi^{out}_{\0} + \phi^{out}_{\1}\sqrt{(2M/R^{\3})}
\left[\frac{1}{\sqrt{(2 M /R^{\3})} r} - arc\tanh{(\sqrt{(2M/R^{\3})} r)}\right] ~~~~ (r > R),
\label{gcg03CDMA}
\end{eqnarray}
where $\phi^{in,out}_{\0,\1}$ are constants to be adjusted in order to match the boundary conditions.
In the Newtonian limit where $A\bb{r}\approx B\bb{r} \approx 1$, the constants match one each other so that $\phi^{in}_{\0,\1} \equiv \phi^{out}_{\0,\1} \phi_{\0,\1}$ and the above solutions reduce to
\begin{equation}
\phi\bb{r} = \phi_{\0} + \frac{\phi_{\1}}{r},
\label{gcg08D}
\end{equation}
which satisfies the suppression of the dependence on $r$, as quoted above, as one is away from the concentration of matter.

We emphasize that any prescription for $m\bb{r}$ is necessarily model dependent, i. e. arbitrary functions for $m\bb{r}$ are equivalent to arbitrary functions for $\phi\bb{r}$ and $m\bb{\phi}$.
This is a fairly general assumption, which is independent of the choice of the equation of state and of the dependence of the variable mass on the scalar field.

To describe the connection among the extensive quantities $\rho$, $m$ and $n$ for an adiabatic system of mass varying particles such as a star (or any stellar object), the relevant thermodynamic equations can be summarized by Eqs.~(\ref{star06})-(\ref{gcg02}).
In fact, in order to get a closed system of equations, one requires a further relationship: the equation of state.
In general this express the pressure in terms of energy density and specific entropy, $p = p\bb{\rho, s}$.
However due to the adiabatic conditions set up, the pressure and its explicit dependence on $r$ is simply given in terms of $\rho\bb{r}$, which can be explicitly obtained from the Tolman-Oppenheimer-Volkoff (TOV) equations for the hydrostatic equilibrium.

Assuming that the equation of state $p = p\bb{\rho\bb{r}}$ leads to a single valued dependence of $\rho$ on the radial coordinate for a spherically symmetric mass distribution of matter, one is left to obtain $\rho\bb{m\bb{r}, n\bb{r}, r}$ which satisfies the system of partial differential equations given by
\begin{equation}
\frac{\mbox{d}\rho}{\mbox{d}r} = \frac{\partial \rho}{\partial m}\frac{\mbox{d} m}{\mbox{d}r} + \frac{\partial \rho}{\partial n}\frac{\mbox{d} n}{\mbox{d}r} + \frac{\partial\rho}{\partial r},
\label{star13}
\end{equation}
and Eqs.~(\ref{star06})-(\ref{gcg02}).
Eliminating $p$ from Eqs.~(\ref{star06})-(\ref{gcg02}), one obtains
\begin{equation}
4 \rho = m \frac{\partial \rho}{\partial m} + 3 n \frac{\partial \rho}{\partial n},
\label{star14}
\end{equation}
and consequently,
\begin{equation}
\rho\bb{m, n} = \kappa m^{\1-\3 \alpha} n^{\1 + \alpha}
\label{star15}
\end{equation}
satisfies Eq.~(\ref{star13}) once setting the variable dependence $\alpha \rightarrow \alpha\bb{r}$.
Using once again Eqs.~(\ref{star06})-(\ref{gcg02}) one obtains $\alpha\bb{r} = p\bb{r}/\rho\bb{r}$ and hence the most general solution for the proposed problem is
\begin{equation}
\rho = \rho\bb{m\bb{r}, n\bb{r}, r} = m^{\1-\3 \frac{p}{\rho}} n^{\1 + \frac{p}{\rho}},
\label{star16}
\end{equation}
where, for simplicity, we have omitted the explicit dependence of $m$, $n$, $p$ and $\rho$ on $r$, and we have adjusted the arbitrary constant $\kappa$ in order to have $\rho = m \, n$ in the non-relativistic limit, and $\rho = n^{\4/\3}$ in the ultra-relativistic limit.

\begin{figure}
\vspace{-.5 cm}
\centerline{\psfig{file=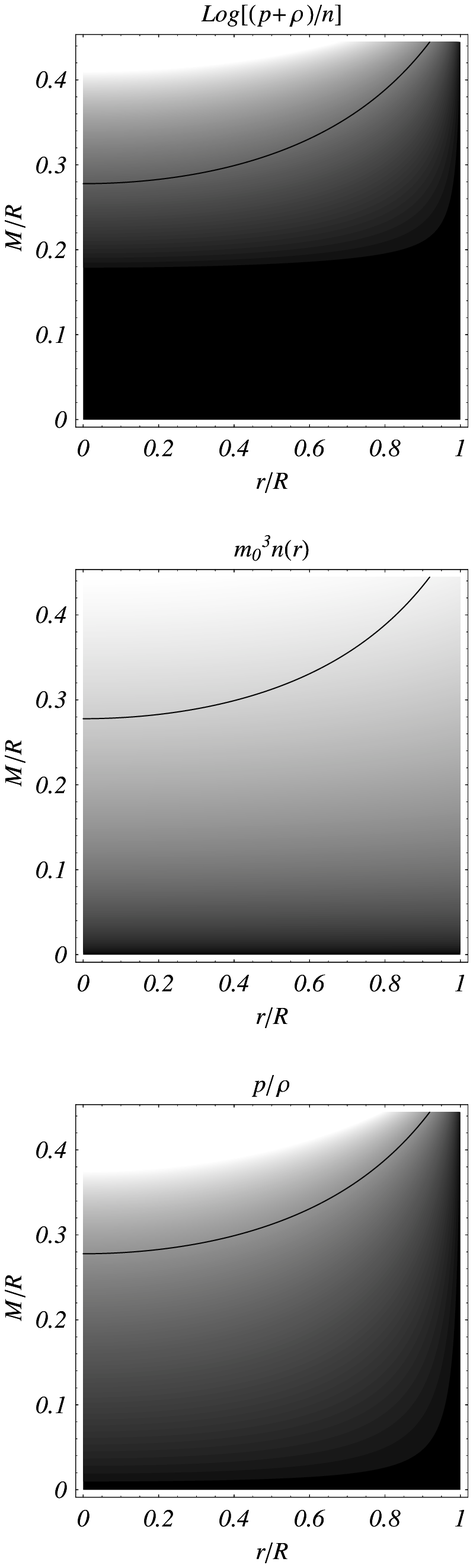}}
\vspace{-.5 cm}
\caption{Relevant thermodynamic quantities for a system composed by particles for which the mass particle is considered as an extensive thermodynamic degree of freedom.
The plots are for the limiting case where the mass of the particle is given by its observable value $m\bb{R} = m_{\0}$.  Notice that the linearly increasing {\em gray level} corresponds to increasing values of the thermodynamic variables, for which we have marked with the boundary values for $M/R = 5/18$, the ``soft'' Buchdahl's limit for which $\rho = 3 p$ at $r = 0$.}
\label{Fstar02}
\end{figure}

Considering a relativistic incompressible fluid, which is actually a rather realistic model for spherically symmetric stellar objects, the energy density is constant out to the surface, i. e. $\rho\bb{r} = \rho\bb{R} = \rho_{\0}$, after which it vanishes.
More specifically, the explicit solution for $p\bb{r}$, and $\rho\bb{r}$ is thus the equation of state.
The equations for the hydrostatic equilibrium \cite{Tol39,Vol39} yields
\begin{equation}
\frac{p\bb{r}}{\rho\bb{r}} = \frac{\left(1 - 2(M/R)(r^{\2}/R^{\2})\right)^{\1/\2} - \left(1 - 2(M/R)\right)^{\1/\2}}{ 3\left(1 - 2(M/R)\right)^{\1/\2} - \left(1 - 2(M/R)(r^{\2}/R^{\2})\right)^{\1/\2}}.
\label{star17}
\end{equation}
for which Newton's constant have been set to unit, $G = 1$, and $M \equiv M\bb{R}$ is the total mass of the compact structure.
As expected, the pressure increases near the core of the star.
The properties of some relevant thermodynamic variables are illustrated in the Fig.~\ref{Fstar02} for the case where their behaviour are constrained by Eq.~(\ref{star16}), in the limit for which the mass of the particle is given by its observable value $m\bb{R} = m_{\0}$.
One computes the ratio between pressure and energy density, $p/\rho$, the density of particles, $n$, dimensionally normalized by $m_{\0}^{\3}$ and the ratio $(p +\rho)/n$ which characterizes a {\em pseudo}-chemical potential.
Indeed, for a star of radius $R$, the central pressure goes to infinity if the mass exceeds $4R/9$, and overpass the radiation pressure, $p = \rho/3$, if the mass exceeds $5R/18$ (solid line in the Fig.~\ref{Fstar02}).

Thus, for a mass beyond the Buchdahl's limit $4R/9$ \cite{Buc59}, general relativity admits no static solutions, that is, a star that shrinks to such size keeps on shrinking, till it turns into a black hole.
This scenario can be modified for astrophysical structures made of particles where mass is constraint by some mass varying mechanism.

Depending on the coupling to the scalar field, consistent with the space-time cosmological curvature, the stability of equilibrium configurations have to be reinterpreted.
Notice that $M$, the total energy inside a radius $R$, includes the rest mass-energy $M_{\0}$, the internal energy of motion $W$ and the (negative) potential self-gravitational energy $U$.
In the relativistic domain, one has:
\begin{equation}
M = 4 \pi \int_{\0}^{\R}{r^{\2} \rho\bb{r}\,\mbox{d}r},
\label{star18A}
\end{equation}
where
\begin{equation}
M_{\0} = 4 \pi \int_{\0}^{\R}{r^{\2} \,m\bb{r} n\bb{r} \sqrt{A\bb{r}}\,\mbox{d}r}.
\label{star18B}
\end{equation}
From Eq.~(\ref{star15}), $M$ reduces to $M_{\0}$ for vanishing binding energy and no motion, that is, if $p = 0$.
The difference between the total rest mass-energy and the total energy corresponds to the positive binding energy $M_{\B}$ which keeps the stellar structure stable,
\begin{equation}
M_{\B} =  M_{\0} - M = 4 \pi \int_{\0}^{\R}r^{\2}{\left[ \rho\bb{r} - m\bb{r} n\bb{r}\sqrt{A\bb{r}}\right]\mbox{d}r}.
\label{star18C}
\end{equation}
where, to evaluate the integration, $n\bb{r}$ is obtained via Eq.~(\ref{star16}) by assuming a constant energy density within the astrophysical structure, and some functional dependence of the particle masses on $r$.

Although it is useful to interpret $M$ as the total energy, the rest energy, $M_{\0}$, is a fundamental quantity in the stability analysis.
Notice that the internal energy of motion, $W$, and the (negative) potential self-gravitational energy, $U$, are particularly relevant only in the Newtonian approximation, where $M_{\B} \approx U + W$.
The binding energy is often referred to as {\em mass defect} and it corresponds to the released energy during the formation of a compact object from an initially rarefied distribution of matter, a typical mechanism for lumping.
When particles are combined into a bound system, an energy corresponding to the mass defect is released in the form of photons, relativistic neutrinos, or gravitational waves.
From the physics of this process, it follows that $M_{\B} > 0$ sets the equilibrium of a static star arising from diffuse matter.

\begin{figure}
\vspace{-.5 cm}
\centerline{\psfig{file=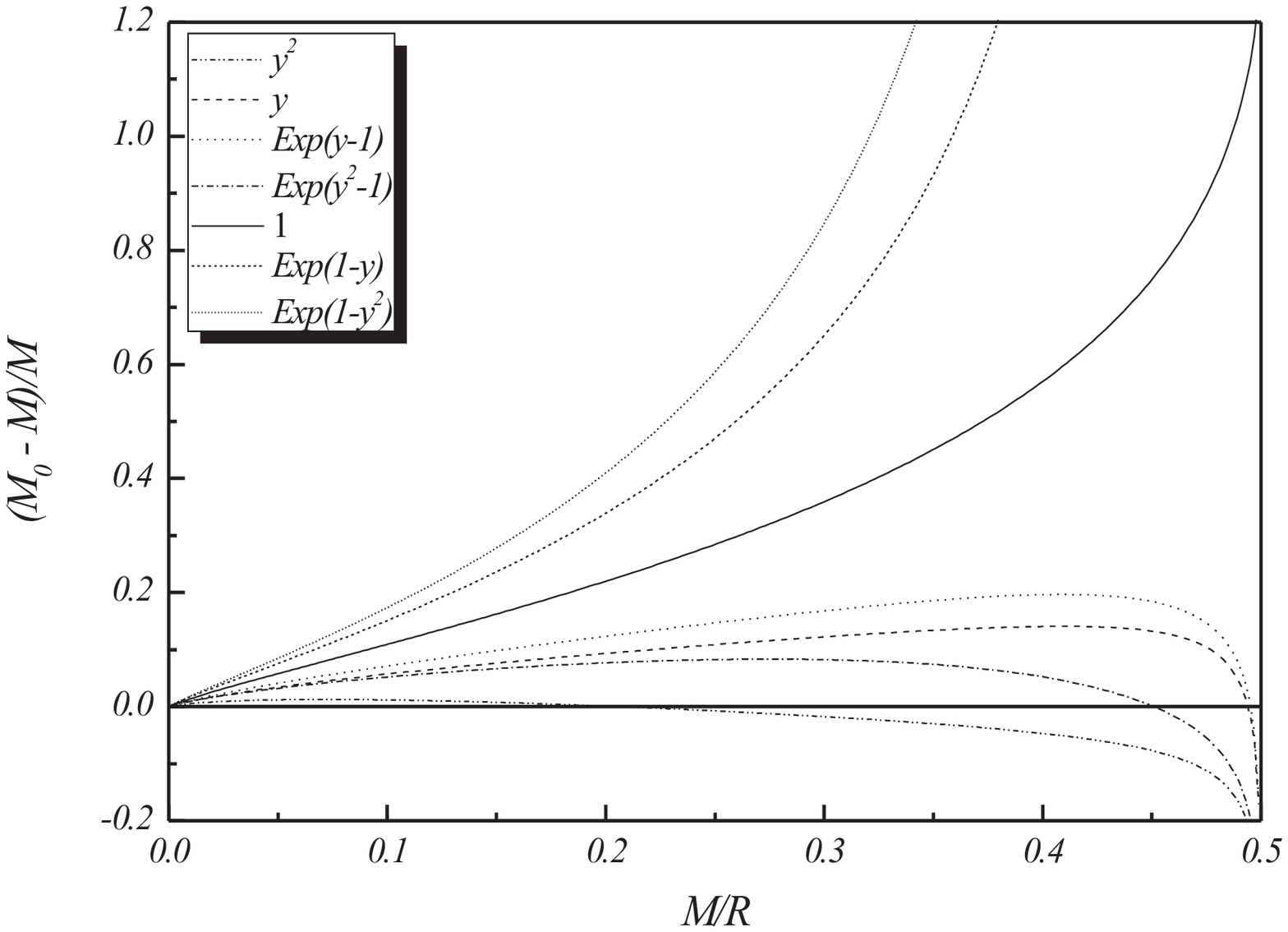, width=14cm}}
\vspace{-.5 cm}
\caption{Mass defect (binding energy) for several mass varying particle functions for compact objects with uniform energy density.
Given that $\phi$ depends on the radial coordinate $r$ for generic classes of curved spaces, one quantifies the modifications due to the present approach by some mass dependencies on $r$ as illustrated in the box, where $\mbox{y} = r/R$.
}
\label{Fstar04}
\end{figure}

In Fig.~\ref{Fstar04} we show how the mass varying mechanism can modify the equilibrium condition based on the mass defect criterium for relativistic stellar objects.
The lumping conditions as prescribed by the total binding energy of the system are modified by the mass varying mechanism.
In the absence of mass varying mechanisms the rest mass-energy $M_{\0}$ is usually written as $M_{\0} = m \, N$, where $N$ is the total number of particles inside the radius $R$ given by
\begin{equation}
N = 4 \pi \int_{\0}^{\R}{r^{\2} n\bb{r} \sqrt{A\bb{r}}\,\mbox{d}r}.
\label{star19}
\end{equation}
The equilibrium curve for this case is described by the solid line in the Fig.~\ref{Fstar04}, for which the binding energy has an upper limit whether one assumes that the pressure at the center cannot goes to infinity, i. e. $M/R < 4/9$ .
It is important to observe that, in case of mass varying particles, the study of the stability in terms of $N$ has to be performed in terms of $M_{\0}$ as inferred from Fig.~\ref{Fstar04}.

Analyzing the mass defect for compact objects composed by particles with masses that increase exponentially inwards to the center reveals that stable configurations are greatly facilitated.
In the case of compact objects with uniform energy density, their structure has a binding energy that increases to infinite as the ratio $M/R$ tends to the Buchdahl's limit.
When these objects reach this situation, they inevitably keep on shrinking, and meeting the conditions to black hole formation.
That is not the case of compact objects with non-uniform energy densities that increase inwards to the center.

On the other hand, for masses that decrease inwards to the center, stable structures can also form up to certain small limiting values of $M/R$.
In opposition to the previous case, the equilibrium conditions are achieved just for smaller values of $M/R$, which restricts the existence of lumps of compact objects with arbitrarily small masses.
The main point is that the coupling with the background scalar field is relevant in determining the conditions for the thermodynamic equilibrium of compact structures.

Thus, the fate of these neutrino lumps, dark matter clusters, or any kind of non-baryonic matter compact structure, depends on the details of the dynamical formation mechanism.
Whether they collapse or not depend on the scalar mediated attractive interaction.
In fact, the role of quintessence potentials in different cosmological scenarios with mass varying neutrinos forming compact structures clearly deserves a deeper analysis.
Since the Higgs sector \cite{Ber09,Ber09C} and the neutrino sector are possibly the only ones where one can couple a new Standard Model (SM) singlet without upsetting the known phenomenology, the replacement of the explicit dependence of the neutrino masses on the scalar field $\phi$ by a direct link to the radial coordinate of the curved space is a quite logical step.
Allowing for the scalar field possibly associated to dark energy to couple with the SM neutrinos and the electroweak interactions may bring important insights into the physics beyond the SM.
The case of cosmological neutrinos, in particular, is a fascinating example where salient questions concerning SM particle phenomenology can be addressed and hopefully better understood.

\begin{acknowledgments}
A. E. B. would like to thank for the financial support from the Brazilian Agencies FAPESP (grant 08/50671-0) and CNPq (grant 300627/2007-6).
\end{acknowledgments}

\pagebreak
\newpage

\end{document}